# A Dual-Core Model for ENSO Diversity: Unifying Model Hierarchies for Realistic Simulations

Jinyu Wang[1,2,3], Xianghui Fang[1,2,3*], Nan Chen[4*], Bo Qin[1,2,3], Mu Mu[1,2,3], Chaopeng Ji[1,2,3]

[1]Department of Atmospheric and Oceanic Sciences & Institute of Atmospheric Sciences, Fudan University, Shanghai, China

[2] Shanghai Key Laboratory of Ocean-Land-Atmosphere Boundary Dynamics and Climate Change, Fudan University, Shanghai, China

[3] Shanghai Frontiers Science Center of Atmosphere-Ocean Interaction, Shanghai, China

[4]Department of Mathematics, University of Wisconsin-Madison, Madison, WI, USA

March, 2025
Dateline

*Corresponding author address*:

  Dr. Xianghui Fang
  Department of Atmospheric and Oceanic Sciences,
  Fudan University, China.
  Email: fangxh@fudan.edu.cn

  Dr. Nan Chen
  Department of Mathematics,
  University of Wisconsin-Madison, Madison, WI, USA
  Email: chennan@math.wisc.edu

**Abstract**

Despite advances in climate modeling, simulating the El Niño-Southern Oscillation (ENSO) remains challenging due to its spatiotemporal diversity and complexity. To address this, we build upon existing model hierarchies to develop a new unified modeling platform, which provides practical, scalable, and accurate tools for advancing ENSO research. Within this framework, we introduce a dual-core ENSO model (DCM) that integrates two widely used ENSO modeling approaches: a linear stochastic model confined to the equator and a nonlinear intermediate model extending off-equator. The stochastic model ensures computational efficiency and statistical accuracy. It captures essential ENSO characteristics and reproduces the observed non-Gaussian statistics. Meanwhile, the nonlinear model assimilates pseudo-observations from the stochastic model while resolving key air-sea interactions, such as feedback balances and spatial patterns of sea surface temperature anomalies (SSTA) during El Niño peaks and improving western-central Pacific SSTA magnitudes and spatial accuracy. The DCM effectively captures the realistic dynamical and statistical features of the ENSO diversity and complexity. Notably, the computational efficiency of the DCM facilitates a rapid generation of extended ENSO datasets, overcoming observational limitations. The outcome facilitates the analysis of long-term variations, advancing our understanding of ENSO and many other climate phenomena.

**Introduction**

As the most significant source of interannual variability on a global scale (Neelin et al. 1998), the El Niño-Southern Oscillation (ENSO) is characterized by its complex spatiotemporal behavior (Capotondi et al. 2015; Timmermann et al. 2018; Cai et al. 2021). This complexity manifests in the temporal irregularity of event duration and frequency, as well as in the spatial diversity of sea surface temperature anomaly (SSTA) patterns. Specifically, El Niño are categorized as either Eastern Pacific (EP) or Central Pacific (CP) events, depending on whether the SSTAs peak in the EP or CP regions, respectively (Kao and Yu 2009; Kug et al. 2009). Over the past several decades, substantial theoretical and observational progress has been made in studying ENSO (Webster and Lukas 1992; McPhaden et al. 1998; Latif et al. 2001; Guilyardi et al. 2016). These efforts have led to the development of a range of models with different levels of complexity, aiming to qualitatively and quantitatively understand the fundamental characteristics of ENSO (Guilyardi et al. 2020). This hierarchy spans from the simplest conceptual models (Suarez and Schopf 1988; Jin 1997; Thual and Dewitte 2023; Izumo et al. 2024) to intermediate complexity models (ICMs; Zebiak and Cane 1987; Kleeman 1993; Dewitte 2000; Zhang et al. 2003) and extends to coupled general circulation models (CGCMs; Guilyardi et al. 2009; Planton et al. 2021; Vaittinada Ayar et al. 2023).

Different types of ENSO models possess distinct strengths and weaknesses. Conceptual or low-order models, which integrate stochastic noise, efficiently simulate realistic time series of regionally averaged SSTA with accurate statistics (Chen et al. 2022; Chen and Fang 2023; Zhao et al. 2024). However, their oversimplification neglects critical processes, such as subsurface temperature changes or detailed ocean-atmosphere interactions, limiting their ability to capture spatially refined ENSO patterns. The intermediate complexity models or ICMs, such as the classic Zebiak-Cane model (Zebiak and Cane 1987), strike a balance: they include more detailed physics than simple models while remaining easier to use and understand than the most complex models. They have been shown to be valuable tools for theoretical studies (Tziperman et al. 1997; van der Vaart et al. 2000; Xie and Jin 2018) and real-time forecasts (Balmaseda et al. 1994; Chen et al. 2004; Liu et al. 2019). Nevertheless, ICMs still struggle to precisely reproduce the statistics of ENSO-related time series (e.g., power spectra) and the diverse spatial patterns of El Niño in the long-term simulation (Chang et al. 1996; Dewitte 2000; Zhang and Gao 2016). In addition, small changes to ICM parameters can often lead to significant differences in the results (Zebiak and Cane 1987; Geng and Jin 2023a,b). The most complex models, called coupled general circulation models or CGCMs, simulate many climate processes across different scales and are theoretically the best tool for studying ENSO dynamics. However, the long-standing mean state biases in CGCMs seriously restrict their performances in representing the observed spatial patterns related to ENSO variability (Magnusson et al. 2013; Taschetto et al. 2014; Freund et al. 2020; Bayr et al. 2024). CGCMs also require vast amounts of computing power, making them inefficient for many research tasks (Wittenberg 2009).

Improving a single model to capture all aspects of ENSO is challenging, as adjustments to one physical process often disrupt others (Mauritsen et al. 2012; Hourdin et al. 2017; Dommenget and Rezny 2018). A more practical solution may lie in combining models to utilize their individual strengths while minimizing weaknesses. To this end, we develop a "multi-core" modeling framework, aiming to treat distinct ENSO model types, each excelling in specific processes, as modular building blocks. This method resembles ocean data assimilation systems (Stammer et al. 2016) in some ways but replaces real-world observations with simulated data from one or a few ENSO models. For instance, a complex submodel can continuously adjust its calculations using simplified data generated by a basic submodel. At the same time, the complex model refines the quality of this simulated data through its own physics, creating a feedback loop that helps reduce model errors (Judd et al. 2008; Mapes and Bacmeister 2012; Kruse et al. 2022). By keeping the overall system at an intermediate level of complexity, this approach balances physical details with computational efficiency, allowing high-quality and realistic long-term simulations.

Building on this framework, we develop a dual-core model (DCM; Fig. 1) to simulate ENSO diversity. The DCM integrates two components: the spatially extended one-dimensional stochastic model of Chen and Fang (2023; CF23), which accurately reproduces equatorial ENSO dynamics and observed statistics, including extreme events, and a two-dimensional intermediate complexity model (ENBOM) enhanced with deep neural networks, inspired by the Zebiak-Cane framework. In this system, the CF23 generates pseudo-observations of SSTA, while the ENBOM uses these synthetic data to extend simulations into off-equatorial regions while incorporating explicit physics. Evaluations demonstrate that the DCM even outperforms the widely used CGCM Community Earth System Model version 2 (CESM2; Capotondi et al. 2020), in simulating realistic ENSO temporal statistics, spatial diversity, and underlying mechanisms.

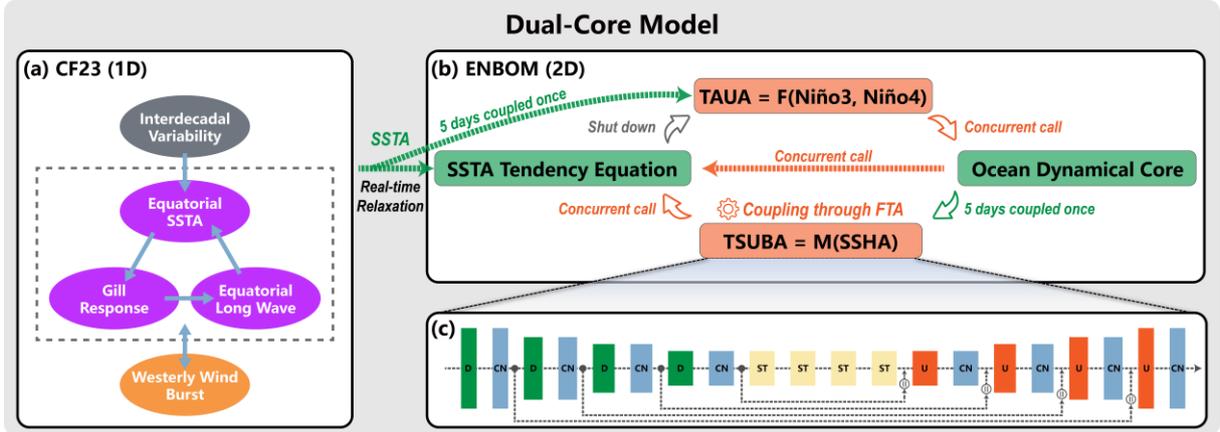

Fig. 1. The schematic illustration of the dual-core model: (a) The one-dimensional and air-sea coupled CF23; (b) The two-dimensional and ocean stand-alone ENBOM that assimilates the pseudo-observation of SSTA from CF23. The TAUA represents the wind stress anomaly in response to Niño 3 and Niño 4 region averaged SSTA of CF23, including zonal and meridional wind stress anomaly TAUXA and TAUYA, respectively; (c) The deep neural parameterization of TSUBA as a function of SSHA. The green, blue, yellow, and orange boxes present the Down-Sampling, ConvNeXt, Swin-Transformer, and Up-Sampling blocks, respectively. More details are introduced in the section of Methods.

**Results**

Given that the CESM2 exhibits relatively high performance in simulating the ENSO among CGCMs (Lee et al. 2021; Planton et al. 2021), we selected its 1200-year preindustrial simulation (CESM2-piControl) from the Coupled Model Intercomparison Project Phase 6 (CMIP6; Eyring et al. 2016) to evaluate long-term internal ENSO variability, which is compared with the DCM developed in this work. These CESM2-piControl data are also utilized for the pre-training phase of the neural TSUBA parameterization scheme. Prior to incorporating the CF23 pseudo-observations, the assimilation procedure within the ENBOM framework was validated against observational data from the Global Ocean Data Assimilation System (GODAS; Behringer and Xue 2004; Supplementary Note 1 and Fig. S1). The subsequent analysis evaluates model performance from three key aspects: (1) fundamental statistical characteristics of ENSO, (2) spatial patterns of physical fields associated with distinct ENSO event types, and (3) oceanic feedback mechanisms governing SSTA evolution.

*Statistics*

The comparison of statistical properties between the SSTA and TSUBA series across the GODAS observation, DCM, and CESM2-piControl includes analyses of power spectral density, seasonal variation, and probability density function (Fig. 2). The DCM realistically reproduces the magnitudes and 3-4-year spectral peaks of the power spectra for observed Niño 4 region (Fig. 2a) and Niño 3 region (Fig. 2d) averaged SSTA (SSTA-N4 and SSTA-N3, respectively), as well as Niño 4 region-averaged TSUBA (TSUBA-N4; Fig. 2g). In contrast, the CESM2-piControl significantly overestimates the power spectra for both SSTA-N4 and TSUBA-N4. Additionally, the DCM aligns well with observations in simulating the seasonal variations of Niño 4 region-averaged SSTA and TSUBA, accurately capturing their magnitudes and the distinct minimum occurring in June (Fig. 2b, h). On the other hand, the CESM2-piControl exhibits excessively large seasonal variability in both SSTA-N4 and TSUBA-N4. Furthermore, the probability density functions of these variables in the DCM are notably closer to observations compared to those in the CESM2-piControl (Fig. 2c, i). The CESM2-piControl, however, displays an unrealistic occurrence of extreme values in the Niño 4 region.

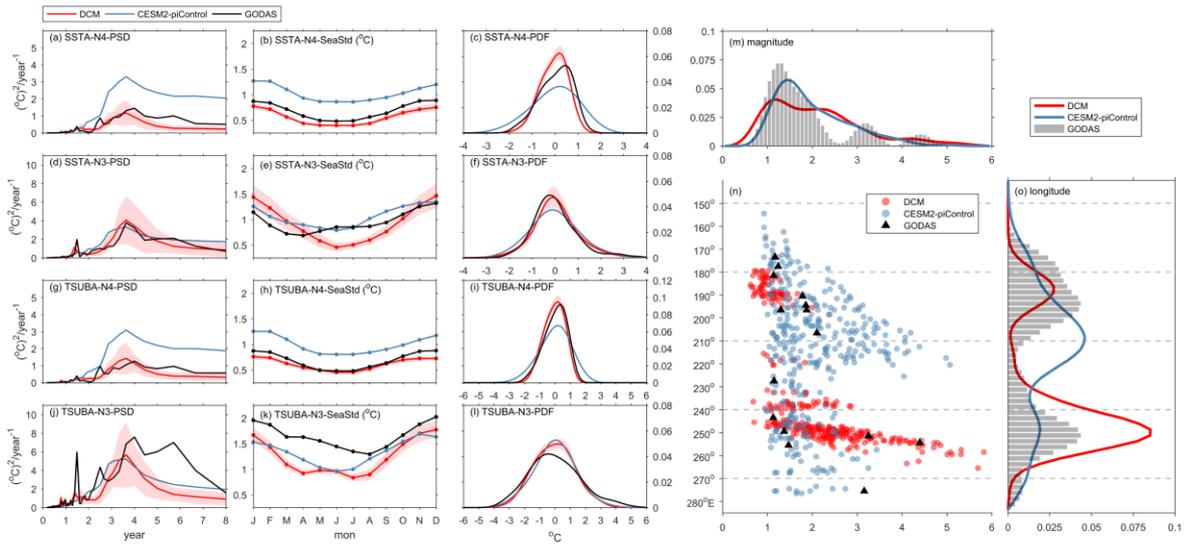

Fig. 2. The power spectral density (PSD; a, d, g, j; unit: (°C)$^2$ /year$^{-1}$), seasonal standard deviation (SeaStd; b, e, h, k; unit: °C), and probability density function (PDF; c, f, i, l) of Niño 4 region averaged SSTA (SSTA-N4), Niño 3 region averaged SSTA (SSTA-N3), Niño 4 region averaged TSUBA (TSUBA-N4), and Niño 3 region averaged TSUBA (TSUBA-N3) for GODAS, DCM, and CESM2-piControl. The solid lines (except the blue line) represent the ensemble mean value for non-overlapping ensemble members with a 40-year duration (one member for GODAS 1981-2020). The pink shading represents one standard deviation of the corresponding values in the DCM to measure the uncertainty. The estimation of power spectral density is based on the 1-2-1 smoothed squares of Fourier transformation of time series. The marginal distribution of (m) magnitude (°C) and (o) longitude (°E) of the maxima of November-January and 5°N-5°S mean SSTA for all the El Niño events in GODAS 1981-2020, DCM 0041-1000, CESM2-piControl 0001-1200. (n) The scatter plot of longitude and magnitude of SSTA maxima.

Given the inherent challenges in classifying EP and CP El Niño events (Johnson 2013; Yu and Kim 2013), we compare the longitude-magnitude distributions of equatorial SSTA maxima during the mature phase of all El Niño events in observations, the DCM, and the CESM2-piControl. While both the DCM and CESM2-piControl exhibit magnitude distributions similar to observations (Fig. 2m), the DCM more accurately reproduces the two prominent peaks in the observed longitude distribution of SSTA maxima (Fig. 2o). In contrast, the CESM2-piControl shows only a single peak in the central Pacific, missing the observed peak in the eastern Pacific. Additionally, the DCM successfully captures the typical emergence of strong SSTA maxima in the eastern Pacific, whereas the CESM2-piControl displays an anomalous concentration of these maxima in the western-central Pacific (Fig. 2n).

*Physical Configurations for Spatial Diversity*

A realistic ENSO model should not only reproduce the statistics of SSTA but also accurately capture the spatial configurations of various physical fields. In terms of SSTA evolution, the observational SSTA patterns are well-matched by the free integration of the DCM (Fig. 3a, b). This includes the diverse representation of CP El Niño (e.g., DCM years 614, 617), extremely strong EP El Niño (e.g., DCM years 582, 586), and consecutive La Niña events (e.g., DCM years 583, 584, 585). As a sharp contrast, the SSTA maxima by the CESM2-piControl typically emerge in the western-central Pacific even for EP El Niño events simulated (Fig. 3c). Furthermore, the DCM successfully captures the observed weaker recharge process of CP El Niño compared to EP El Niño (Supplementary Fig. S2), where CP El Niño tends to transition into a subsequent neutral or weak warm state rather than a cold one (Priya et al. 2024).

The relationships between SSTA-TAUXA, SSTA-SSHA, and SSHA-TSUBA for different types of composite ENSO events are also examined. Consistent with the longitude-magnitude distribution of SSTA maxima, the SSTA centers of EP and CP El Niño events in the DCM are clearly distinguishable (Fig. 3e1, e2) and align well with observations (Fig. 3d1, d2). In contrast, the CESM2-piControl shows similar spatial patterns for EP and CP El Niño, characterized by relatively uniform warming across the entire equatorial Pacific (Fig. 3f1, f2). Additionally, the DCM corrects the westward shift of the SSTA centers in the leading empirical orthogonal functions (EOFs) seen in the CESM2-piControl (Supplementary Fig. S3, S4).

Regarding the zonal wind stress response to SSTA during ENSO events, the DCM reproduces the observed pattern, with the response center confined to the west of 210°E and south of the equator (McGregor et al. 2012; Stuecker et al. 2013; Gong and Li 2024). In contrast, the CESM2-piControl exhibits zonal wind stress anomalies with high values extending eastward beyond 210°E into the eastern Pacific (e.g., Fig. 3f1). This bias in atmospheric response leads to an eastward shift of SSHA, creating an unrealistic configuration where SSHA is shifted eastward while SSTA remains shifted westward. In the DCM, however, the centers of SSHA and SSTA are more closely aligned for both EP and CP El Niño events (Fig. 3e4, e5).

The DCM also provides accurate spatial depictions of TSUBA. For example, the maxima of TSUBA associated with CP El Niño in both observations and the DCM are located in the eastern Pacific but away from the eastern coast (Fig. 3d5, e5). While the standard deviations of TSUBA in the DCM are slightly weaker than in observations, the model successfully captures the observed large-value regions around 10°N (Supplementary Fig. S5). Notably, although SSHA and TSUBA share the same sign over most of the equatorial Pacific, the zero-value point of SSHA on the equator does not precisely coincide with zero TSUBA in observations, the DCM, or the CESM2-piControl. This discrepancy, overlooked by the simplified parameterization schemes of TSUBA in previous ZC-type models (Yuan et al. 2020), highlights the complexity of these interactions.

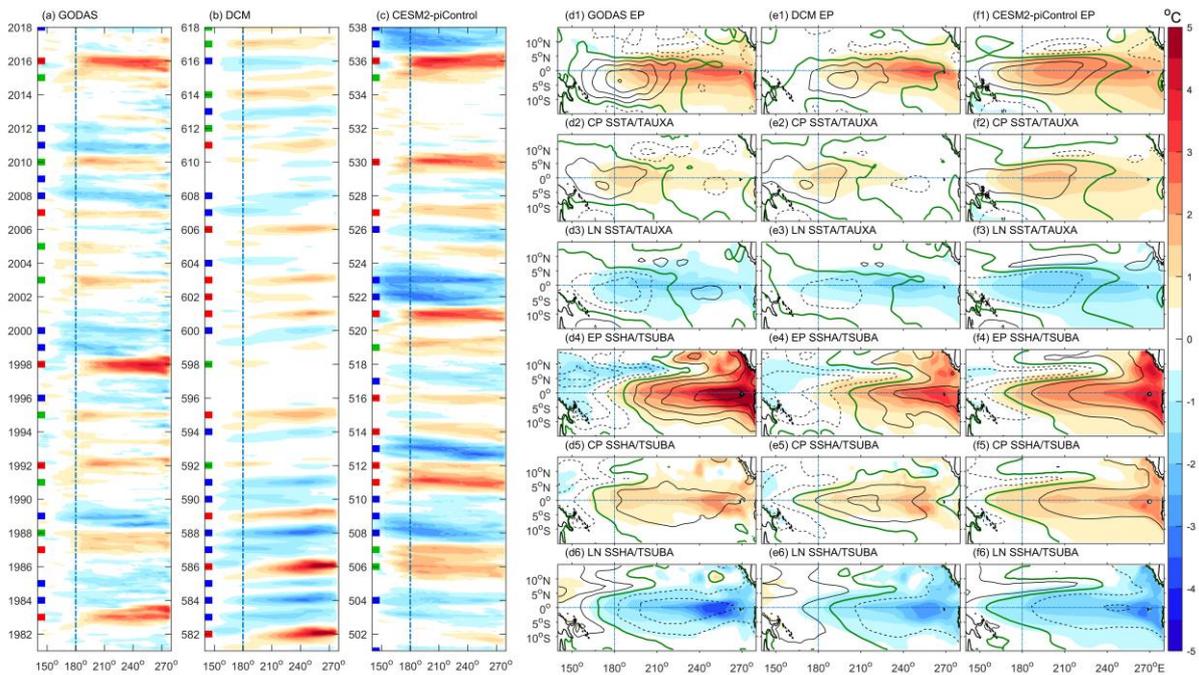

Fig. 3. The time evolution of 5°N-5°S mean SSTA (shading) for GODAS (a), DCM (b), and CESM2-piControl (c) with a shading interval of 0.5°C. The red, green, and blue blocks denote the EP El Niño (EP), CP El Niño (CP), and La Niña (LN) events, respectively. The combined composite spatial patterns of November-December-January mean SSTA (shading)-TAUXA (contour) are in d1-f1, d2-f2, d3-f3 and SSHA (contour)-TSUBA (shading) are in d4-f4, d5-f5, d6-f6 for EP, CP, and LN events, respectively. The shading interval is 0.5°C for both SSTA and TSUBA. The solid black, solid green, and dashed black contour lines denote the positive, zero, and negative values, respectively. The equidistant contour interval is 0.04m for SSHA and 0.01 N m$^{-2}$ for TAUXA. The ENSO events range from 1981-2020, 0041-1000, 0001-1200 for GODAS, DCM, and CESM2-piControl, respectively. Here, the multi-year 2014/15 and 2015/16 El Niño are identified as two events (Song and Yu 2019).

*Physical Mechanism*

Next, we aim to understand whether the DCM can capture the main mechanisms of SSTA's evolution. The SSTA tendency equation can be decomposed into three main feedback terms to conduct a heat budget analysis (Wang 2001; Kug et al. 2010; Capotondi 2013):

$$\frac{\partial T}{\partial t} = TEND = ZAFK + THFK + RESQ \qquad (1)$$

$$ZAFK = -u\frac{\partial \bar{T}}{\partial x} \qquad (2)$$

$$THFK = -M(\bar{w})\left(\frac{\partial T}{\partial z}\right) = -\gamma M(\bar{w})\frac{T - T_{sub}}{H_m} \qquad (3)$$

where $TEND$, $ZAFK$, $THFK$, and $RESQ$ are total tendency, zonal advective feedback, thermocline feedback, and residual terms, respectively. $u$, $T_{sub}, \bar{T}$ and $\bar{w}$ stands for the zonal current anomaly, subsurface temperature anomaly, climatological mean values of SST, and upwelling velocity. $H_m$ is the constant mixed layer depth (50 m). $\gamma$ denotes the thermocline feedback strength or mixed ratio of TSUBA and varies along the equator (Supplementary Note S2). $M(\bar{w})$ is the rectified linear unit function to take the mean (upwelling) Ekman velocity as positive. The residual term accounts for most of the model uncertainties in the ZC-type SSTA tendency equation, such as damping, diffusion, and the missing diabatic heating. Specifically, we assume that the thermocline feedback terms in observation and CESM2-piControl follow the same expression in the DCM to ensure a fair comparison (Fig. 4). It is essential to ensure that the physical balance of the SSTA tendency equation in the external model ENBOM is not invaded too much by the pseudo-observation from the internal model CF23.

In observations, the zonal advective feedback (Fig. 4a1, d1) and thermocline feedback (Fig.

4a2, d2) are identified as the two dominant positive feedback mechanisms throughout the development of both EP and CP El Niño events. The DCM successfully captures the leading role of the linear dynamic oceanic feedback. For instance, the residual term in the DCM primarily acts as negative feedback, suppressing SSTA during the mature phases of both EP and CP El Niño events (Fig. 4b3, e3). The spatiotemporal structure of oceanic feedback associated with La Niña in the DCM closely resembles that of EP El Niño (Supplementary Fig. S6). In CESM2-piControl, thermocline feedback west of 240°E is weaker than in observations for both types of El Niño (Fig. 4c2, f2), reflecting a deeper thermocline in the central Pacific and weaker equatorial upwelling (Pang et al. 2025). Additionally, the residual terms consistently contribute positive feedback during the development years of EP El Niño in CESM2-piControl (Fig. 4c3). These issues suggest that the SSTA evolution in CESM2 cannot be adequately approximated by the ZC-type tendency equation.

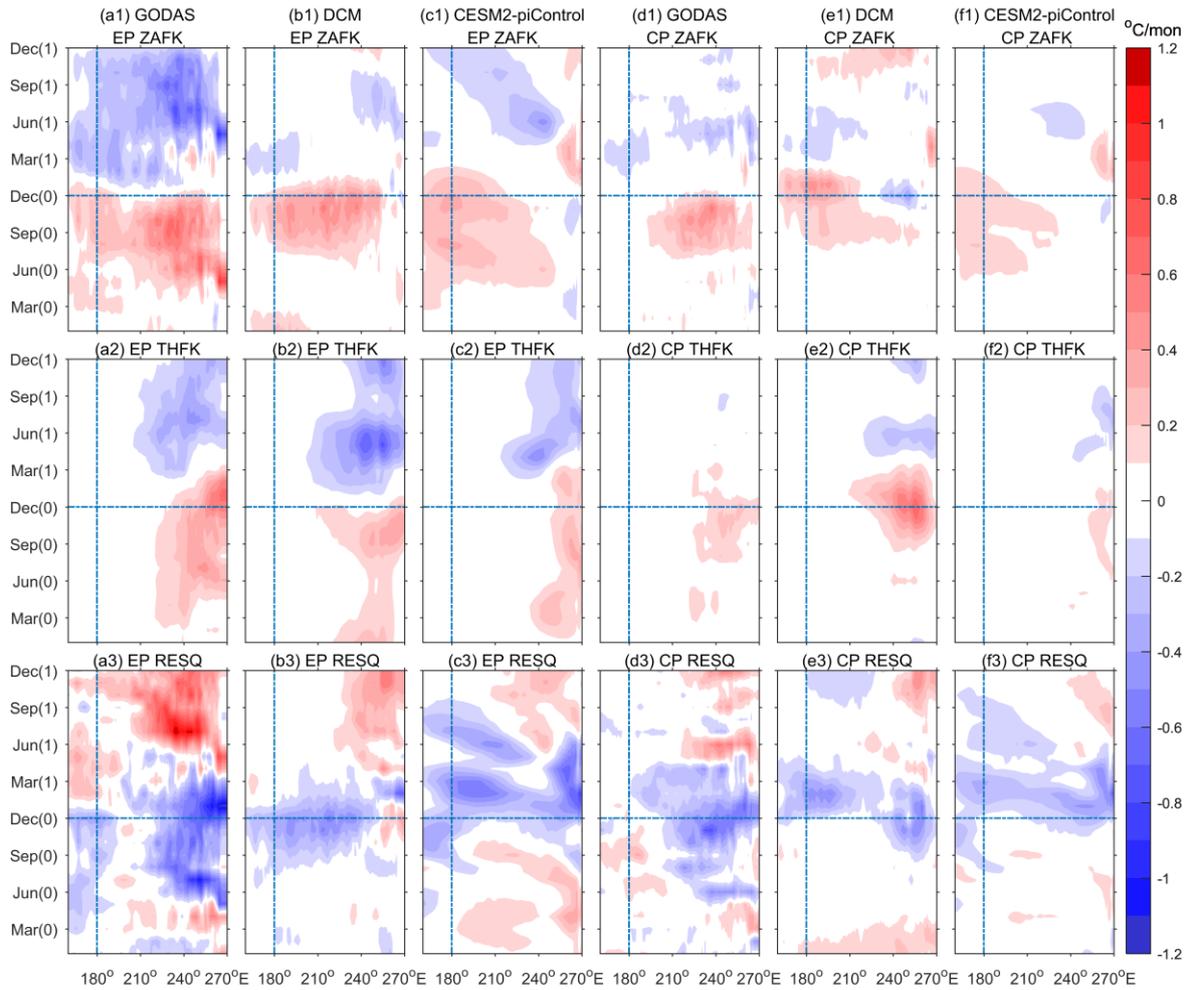

Fig. 4. The composite time evolution of 5°N-5°S mean heat budget for the development year (0) and the subsequent year (1) of events of EP El Niño (the first three columns) and CP El Niño (the last three columns). The rows of figures, arranged from top to bottom, denote the zonal advective feedback (ZAFK), thermocline feedback (THFK), and residual terms (RESQ). The titles of the first row denote GODAS, DCM, and CESM2-piControl, respectively. The shading interval of these three terms is 0.1°C per month (°C mon$^{-1}$). The labels of the vertical axis stand for the end of calendar months.

As indicated by the advective-reflective oscillator theory (Picaut et al. 1997), zonal advective feedback in the DCM and CESM2-piControl are primarily confined to the western-central Pacific. This can be attributed to the strong zonal gradient of mean SST near the eastern edge of the warm pool and the pronounced zonal current anomalies driven by local wind stress anomalies. In contrast to simulations, extensive zonal advective feedback appears in the eastern Pacific in observations (Fig. 4a1, d1), likely due to strong zonal current anomalies in that region. However, the underestimation of eastern Pacific zonal currents remains a common issue in CGCMs (Lin et al. 2023) and even in simplified shallow water models forced by observational winds (Wang et al. 2024). Nevertheless, in the DCM, zonal advective feedback can extend further into the eastern Pacific for EP El Niño compared to CP El Niño.

**Discussion**

This study demonstrates the ENSO simulation using a DCM, which integrates the advantages of different types of ENSO models while maintaining constrained complexity. Our primary goal is to depict the spatiotemporal characteristics of ENSO as realistically as possible, using observations and CESM2-piControl simulations as benchmarks. One of the most notable results is that our DCM is more effective in reproducing the marginal longitude distribution of SSTA maxima during the peaks of El Niño events. Secondly, the power spectral density and seasonal variations of Niño 4-averaged SSTA, as well as the TSUBA series, are more realistic in the DCM. Furthermore, our simulations significantly reduce biases in the spatial patterns and configurations of SSTA, TAUXA, SSHA, and TSUBA associated with both EP and CP El Niño events in CESM2. Specifically, we address the overly eastward biases of TAUXA and SSHA, as well as the overly westward bias of SSTA. Lastly, the zonal advective feedback in the western-central Pacific and the thermocline feedback in the eastern Pacific remain the dominant positive feedback mechanisms during the development years of ENSO events.

Our study confirms that the dual-core model has the potential to transform an unfavorable dilemma, where higher-resolution models, while resolving more physics, make performance optimization more challenging, into a favorable condition for ENSO modeling. This is achieved by incorporating a statistically accurate simple model to aid in optimization while preserving detailed physics. With its superior performance in depicting ENSO, the long-term simulation datasets from the DCM could serve as valuable resources for pre-training artificial intelligence proxy models in simulating and predicting ENSO (Ham et al. 2019; Qin et al. 2024) as well as facilitating the analysis of extreme events and long-term variations. Furthermore, the general framework developed here allows the design of a multi-core network with specific connections for integrating more than two climate models, which has the potential to incorporate additional desirable features and processes that interact with ENSO. Overall, simple and complex climate models could coexist within a flexible platform designed to study the mechanisms and predictability of ENSO.

## Methods

### Data

The observational data for sea surface temperature, sea surface height, currents, vertical velocity, and wind stresses are sourced from the following datasets: monthly averaged Global Ocean Data Assimilation System (GODAS; Behringer and Xue 2004) spanning 1980 to 2023, monthly averaged Simple Ocean Data Assimilation version 2.2.4 (SODA2; Giese and Ray 2011) covering 1901 to 2008, and 5-day averaged Simple Ocean Data Assimilation version 3.12.2 (SODA3; Carton et al. 2018) from 1981 to 2016. For training deep neural networks and evaluating simulation performance, a 1200-year CESM2-piControl simulation dataset is utilized. Climatological states are defined as the multi-year monthly mean states for specific periods: 1991–2020 for GODAS, 1971–2000 for SODA2, 1981–2010 for SODA3, and years 1–1200 for CESM2-piControl. Anomalies in CESM2-piControl are calculated as departures from these climatological states, while anomalies in the other datasets are derived as deviations from the multi-year monthly mean states of detrended primitive data to account for significant global warming. Subsurface temperature anomalies (TSUBA) are defined as temperature anomalies at a depth of 50 meters, representing the constant mixed layer thickness. The Earth

Topography 2022 dataset (NOAA NCEI 2022) is used to determine land boundaries. Additionally, equatorial baroclinic normal modes of the ocean are computed using the climatological mean density of seawater from the World Ocean Atlas 2018 (Locarnini et al. 2019).

Referring to Kug et al. (2009), different types of ENSO events in observations and model simulations are classified based on the following criteria: For El Niño (La Niña) events, the 3-month running mean values of SSTA in either the Niño 3 or Niño 4 regions must exceed (or fall below) 0.7 (-0.7) standard deviations, persist for at least five consecutive months, and span the end of the year. If the November-December-January or December-January-February mean SSTA in the Niño 4 region (SSTA-N4) is greater than that in the Niño 3 region (SSTA-N3), the event is classified as a CP El Niño. Otherwise, it is identified as an EP El Niño event.

*Models*

The Dual-Core Model (DCM) integrates two ENSO models. The simpler model, CF23, focuses solely on ENSO-related fields along the equator of the tropical Pacific Ocean, with a zonal resolution of 2.8125° longitude (56 grid points spanning 126°E to 280.6875°E). It comprises three key components: (1) an interannual air-sea coupled linear dynamic system, (2) an external independent stochastic process representing the background strength of the interdecadal Walker circulation, and (3) westerly wind bursts parameterized using multiplicative stochastic noise that depends exclusively on sea surface temperature anomalies in the Niño 4 region (SSTA-N4). Here, the self-varying background interdecadal variability modulates the occurrence and intensity of CP and EP El Niño events (Hu and Fedorov 2018; Stevenson et al. 2019; Zhao and Fedorov 2020) while the wind burst parameterization functions as a stochastic perturbed parameterization tendencies scheme, enhancing ENSO simulation accuracy, particularly in reproducing power spectra (Christensen et al. 2017; Strømmen et al. 2018).

The more complex model, ENBOM, is an anomalous model covering the tropical Pacific and Indian Oceans between 30°N and 30°S, with a horizontal resolution of 1° in both longitude and latitude (288 × 64 grid points). Following Dewitte (2000) and Zhao et al. (2021), SSHA, zonal current anomalies, and meridional current anomalies are computed using a linearized multi-mode shallow-water model that incorporates the three gravest climatological baroclinic modes, which vary along the equator. Additionally, vertical Ekman-pumping velocity anomalies at the base of the mixed layer (50 m depth) are estimated using a diagnostic Ekman shear layer model (Wang and Fang 1996). Based on the GODAS dataset, the statistical wind stress response to SSTA is modeled through monthly varying linear bivariate regression onto SSTA-N4 and SSTA-N3 ($Niño\ 4\ (t)$ and $Niño\ 3\ (t)$):

$$TAUA = W_C(x, y, cmon) Niño\ 4\ (t) + W_E(x, y, cmon) Niño\ 3(t) \qquad (4)$$

where $TAUA$ stands for the wind stress anomalies, $W_C(x, y, cmon)$ and $W_E(x, y, cmon)$ are the bivariate regression coefficients for 12 calendar months. The TSUBA in the module of the SSTA tendency equation is only determined by SSHA, following the physical assumption of the ZC-type model. The time-frequency of wind stress response to SSTA and TSUBA response to SSHA is set to once every 5 days, remaining unchanged during this period.

Here, transformer-type deep neural networks are employed to parameterize TSUBA (Supplementary Note S3). This approach is advantageous because neural networks can mitigate artificial uncertainties associated with the use of finite elementary functions in traditional physical parameterizations (Rasp et al. 2018; Yuval and O'Gorman 2020; Kochkov et al. 2024). For example, in ICMs, TSUBA has been locally parameterized using differentiable hyperbolic tangent activation functions of surface potential heights (Jin 1996; Yuan et al. 2020), which are functionally equivalent to single-layer neural networks. To strengthen the connection between the model and observational data, the shallow-water-type model is driven by 3.3-month (100-day) low-pass filtered wind stress anomalies from the SODA2 and SODA3 datasets, while the GODAS dataset is reserved for calibrating and tuning the entire ENBOM. Simultaneously, the low-pass filtered TSUBA data are used for the "online" training of deep neural networks, which consist of 40 million parameters. To prevent overfitting due to the limited volume of observational data, the 1200-year, 3.3-month low-pass filtered CESM2-piControl dataset, including SSHA and TSUBA, is also utilized for pre-training. To eliminate mesoscale noise, the output of the TSUBA parameterization is reconstructed using the first 96 empirical orthogonal functions (EOFs) of TSUBA from the GODAS dataset, which collectively explain 92.96% of the total variance. Furthermore, the deep-learning parameterization is implemented using the PyTorch framework, while the other models are coded in Fortran. The Fortran-Torch-Adapter facilitates the integration of PyTorch libraries with Fortran programs (Mu et al. 2023).

*Dual-Core Simulation*

Although the ENBOM is based on first-principle physical assumptions of ENSO, it is only capable of generating a regular interannual air-sea coupled oscillation and fails to accurately simulate some key features of ENSO, such as the seasonal variations (Supplementary Fig. S7). Adding to this challenge, ENBOM remains a highly complex model with potential limitations. The parameters of this coupled system, integrated with deep nonlinear neural networks, are excessively large (40 million), making purposeful and effective tuning impractical, particularly for quantitative analysis of low-order dynamic systems (Wittenberg and Anderson 1998; Dijkstra 2019). Given that CF23 has already been extensively tuned against the GODAS dataset, a one-way communication approach is adopted here. Specifically, the two models run simultaneously, sharing the same timeline. The ENBOM serves as the external "shell" of CF23, estimating the states of interannual anomalous physical fields constrained by real-time pseudo-observations of CF23. It worth noting that the computational cost of ENBOM is tens of times more expensive than that of the simpler CF23, and the running time of CF23 is negligible during the assimilation.

Regarding the methods of assimilating the SSTA-determined information, the ENBOM is designed to operate under the air-sea uncoupling configuration. Its ocean component is forced by its atmospheric component in response to the SSTA-N4 and SSTA-N3 of CF23 over a 5-day period, without any changes during this time. Similarly, the neural networks of TSUBA operate once every 5 days. What's more, the SSTA of ENBOM is concurrently relaxed to the reconstructed pseudo-observation from CF23 with a time scale of 60 days. The linear Newtonian relaxation or nudging is selected for the heuristic assimilation of SSTA:

$$\frac{\partial T}{\partial t} = F(T,t) - \frac{T - T_{POBS}}{\tau} \quad (5)$$

where $\frac{\partial T}{\partial t}$, $F(T,t)$, and $\left(-\frac{T-T_{POBS}}{\tau}\right)$ are the total tendency of SSTA, the physical feedback term belonging to the ENBOM, and the nudging term, respectively. The SSTA-N4 and SSTA-N3 of CF23 are used to reconstruct pseudo-observation of SSTA over the tropical Pacific Ocean based on the 3.3-month low-passed data of GODAS (Chen et al. 2022):

$$T_{POBS} = R_C(x,y)\,Niño\,4(t) + R_E(x,y)Niño\,3(t) \qquad (6)$$

where $T_{POBS}$ stands for the pseudo-observation of SSTA, $R_C(x,y)$ and $R_E(x,y)$ are the regression patterns associated with SSTA-N4 and SSTA-N3 of CF23. Chen et al. (2022) also emphasized that this bivariate linear regression of SSTA can explain the vast majority of SSTA variances along the equator. Specifically, the SSTA-N4 and SSTA-N3 remain rigorously unchanged before and after the reconstruction. The synchronized integration of two ENSO models persisted for 1,000 years. The monthly mean outputs for the last 960 years are selected for analysis.

**Data Availability:**

The GODAS data are obtained from https://psl.noaa.gov/data/gridded/data.godas.html. The SODA2 and SODA3 data are both openly available at https://www2.atmos.umd.edu/~ocean/. The CESM2-piControl simulation data are obtained from https://esgf-node.llnl.gov/projects/cmip6/ The ETOPO 2022 data are freely available at https://www.ncei.noaa.gov/products/etopo-global-reliefmodel. The WOA18 data are from https://www.ncei.noaa.gov/products/world-ocean-atlas.

**Code Availability:**

The source code of the DCM is available at https://github.com/BrunoQin/DCM. The download and the installation of Fortran-Torch-Adapter are accessible at https://github.com/luc99hen/FTA.

**Acknowledgments:**

The research of Jinyu Wang, Xianghui Fang, Mu Mu, Bo Qin, and Chaopeng Ji is supported by the National Natural Science Foundation of China (Grant Nos. 42288101 and 42192564), the Ministry of Science and Technology of the People's Republic of China (Grant No. 2020YFA0608802). The research of Nan Chen is funded by the Office of Naval Research N00014-24-1-2244 and the Army Research Office W911NF-23-1-0118.

**Author contributions statement:**

J. W., X. F., N. C., and B. Q. designed the project. J.W., B.Q., N.C., and X.F. contributed the model codes. J.W., and B.Q. conducted model experiments. J.W., X.F., N.C., and B.Q. wrote the manuscript. J.W., X.F., N.C., B.Q., M.M., and J.C. discussed the results and reviewed the manuscript.

**Competing interests:**

The authors declare no competing interests.

Supporting Information for

# A Dual-Core Model for ENSO Diversity: Unifying Model Hierarchies for Realistic Simulations

Jinyu Wang[1,2,3], Xianghui Fang[1,2,3*], Nan Chen[4*], Bo Qin[1,2,3], Mu Mu[1,2,3], Chaopeng Ji[1,2,3]

[1]Department of Atmospheric and Oceanic Sciences & Institute of Atmospheric Sciences, Fudan University, Shanghai, China

[2] Shanghai Key Laboratory of Ocean-Land-Atmosphere Boundary Dynamics and Climate Change, Fudan University, Shanghai, China

[3] Shanghai Frontiers Science Center of Atmosphere-Ocean Interaction, Shanghai, China

[4]Department of Mathematics, University of Wisconsin-Madison, Madison, WI, USA

## Contents of this file

**Supplementary Note S1. Preliminary Validation of the Assimilation Method Using GODAS Dataset**

Before incorporating the pseudo-observation of SSTA generated by the simple reduced-order ENSO model CF23, we conducted three preliminary experiments to validate the performance of the assimilation method using the Global Ocean Data Assimilation System (GODAS) dataset (Supplementary Fig. S1). First, the ocean component of ENBOM was directly forced by wind stress anomalies from the 1981–2020 GODAS dataset. Second, it was driven solely by wind stress anomalies reconstructed from SSTA-N4 and SSTA-N3 derived from GODAS. Third, building on the second experiment, the SSTA of ENBOM was nudged toward the reconstructed pseudo-observation of SSTA from GODAS using a 60-day nudging time scale. These experiments are referred to as FULL-FORCE, REG-FORCE, and NUDGE, respectively.

The results show that the SSTA-determined atmospheric component effectively filters out higher-frequency atmospheric "noise" in equatorial wind stress anomalies (Supplementary Fig. S1c). However, the absence of this atmospheric noise improves the correlation coefficient associated with SSTA-N3 (Supplementary Fig. S1a, b). Meanwhile, the correlation coefficients for SSTA-N4 remain around 0.9 in both the FULL-FORCE and REG-FORCE runs. These findings indicate that the current ENBOM demonstrates enhanced skill in reproducing ENSO-related time series when it assimilates interannual SSTA-determined information. Notably, in the NUDGE run, the correlation coefficients for SSTA-N4 and SSTA-N3 improve to approximately 0.95 compared to the REG-FORCE run, while other series remain unchanged. These results provide logical support for the dual-core simulation framework, where ENBOM assimilates only the SSTA-determined pseudo-observations provided by CF23.

**Supplementary Note S2. The SSTA tendency equation**

As outlined in the main text, the SSTA tendency equation in the ICM follows the expressions (Eqs. S1-S2) given by Zebiak and Cane (1987) :

$$\frac{\partial T}{\partial t} = -u\frac{\partial \bar{T}}{\partial x} - \bar{u}\frac{\partial T}{\partial x} - u\frac{\partial T}{\partial x} - v\frac{\partial \bar{T}}{\partial y} - \bar{v}\frac{\partial T}{\partial y} - v\frac{\partial T}{\partial y}$$
$$-[M(\bar{w}+w) - M(w)]\frac{\partial \bar{T}}{\partial z}$$
$$-M(\bar{w})\frac{\partial T}{\partial z} - [M(\bar{w}+w) - M(w)]\frac{\partial T}{\partial z} + K\Delta T - \alpha T \quad (S1)$$

$$\frac{\partial T}{\partial z} = \gamma\frac{T - T_{sub}}{H_m} = \frac{T - [(1-\gamma)T + \gamma T_{sub}]}{H_m} \quad (S2)$$

where $(u, v, w, T)$ and $(\bar{u}, \bar{v}, \bar{w}, \bar{T})$ are anomalous and climatological values for zonal current, meridional current, vertical velocity at the bottom of the mixed layer, and sea water temperature, respectively. The zonal profile of thermocline feedback strength $\gamma$ (or mixed ratio of subsurface temperature anomaly) has a maximum value of 0.75 (Supplementary Fig. S8). $K$ and $\alpha$ stand for the horizontal Laplacian diffusion coefficient (3000 m$^2$ s$^{-1}$) and damping coefficient (100 day$^{-1}$) for SSTA. The climatological states of horizontal currents, sea surface temperature, and subsurface temperature (at 50 m depth) are directly assigned using the 1991–2020 mean states from the GODAS dataset. Following Wang and Fang (1996), the vertical velocities at the base of the mixed layer are primarily determined by the convergence of the Ekman shear layer. Consequently, the climatological vertical velocities are derived exclusively from an Ekman shear layer model, which is driven by the 1991–2020 mean wind stresses from the GODAS dataset.

**Supplementary Note S3. The transformer neural networks of TSUBA parameterization**

We employ advanced neural networks to develop a purely data-driven parameterization for TSUBA, with its structure illustrated in Fig. 1c. This model takes SSHA as input and generates the corresponding TSUBA. The mean square error is used as the loss function to train the model. Similar to the U-Net architecture (Ronneberger et al. 2015), the network for this parameterization follows a typical encoder-decoder design, incorporating various functional neural modules. Both the encoder and decoder consist of alternating Down-Sampling (green boxes) and Up-Sampling blocks (red boxes), as well as ConvNeXt blocks (blue boxes) (Liu et al. 2022). The Down-Sampling and Up-Sampling blocks compress or restore gridded data, gradually transforming the receptive field of convolution operators. The ConvNeXt blocks are designed to capture localized latent spatial features within the data, while also enhancing the model's capacity, nonlinearity, and diversity in generating TSUBA. Between the encoder and decoder, four Swin-Transformer blocks (yellow boxes) (Liu et al. 2021) are utilized to learn global characteristics, enabling accurate mapping between SSHA and TSUBA across different ENSO phases. Additionally, skip connections are implemented to bridge the encoder and decoder. These connections propagate and cascade information from shallow to deep layers, ensuring the full utilization of multi-layer features and mitigating the issue of gradient vanishing during deep network training (He et al. 2016).

**Supplementary Table S1. The list of acronyms with their units and descriptions**

| Acronym | Unit | Description |
|---|---|---|
| ENBOM | | intermediate complexity model of ENSO with deep neural networks of subsurface temperature parameterization, named by the equatorial neural baroclinic ocean-determined model |
| CF23 | | Chen-Fang 2023 stochastic reduced-order model of ENSO |
| FULL-FORCE | | Forced by full wind stress anomalies from observation |
| REG-FORCE | | Forced by the reconstructed wind stress anomalies based on the SSTA-N4 and SSTA-N3 observations |
| NUDGE | | Nudged by the reconstructed SSTA of observation with a nudging scale of 60 days, and forced by the reconstructed wind stress anomalies based on the SSTA-N4 and SSTA-N3 of observation |
| POBS | °C | Pseudo-observation of SSTA from CF23 |
| SSTA-N4 | °C | Niño 4 region averaged sea surface temperature anomaly |
| SSTA-N3 | °C | Niño 3 region averaged sea surface temperature anomaly |
| TSUBA-N4 | °C | Niño 4 region averaged subsurface temperature anomaly (at the depth of 50m) |
| TSUBA-N3 | °C | Niño 3 region averaged subsurface temperature anomaly |
| TAUXA-N4 | N m$^{-2}$ | Niño 4 region averaged zonal wind stress anomaly |
| SSHA-N3 | m | Niño 3 region averaged sea surface height anomaly |
| TEND | °C mon$^{-1}$ | The total tendency of SSTA |
| ZAFK | °C mon$^{-1}$ | Zonal advective feedback term |
| THFK | °C mon$^{-1}$ | Thermocline feedback term |
| RESQ | °C mon$^{-1}$ | Residual term |

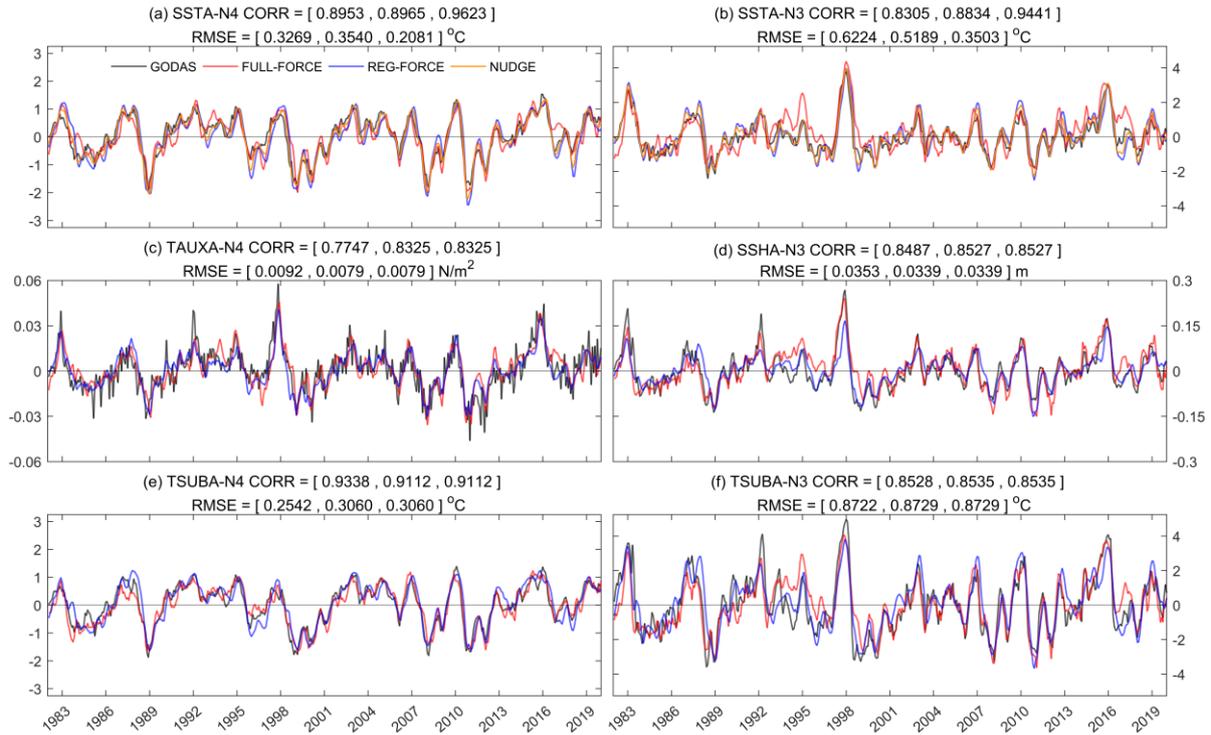

**Supplementary Fig. S1.** The time series of (a) Niño 4 region averaged SSTA (SSTA-N4; °C), (b) Niño 3 region averaged SSTA (SSTA-N3; °C), (c) Niño 4 region averaged zonal wind stress anomalies (TAUXA-N4; N m$^{-2}$), (d) Niño 3 region averaged sea surface height anomalies (SSHA-N3; °C), (e) Niño 4 region averaged TSUBA (TSUBA-N4; °C), and (f) Niño 3 region averaged TSUBA (TSUBA-N3). The solid black, red, blue, and orange lines represent the GODAS observation, FULL-FORCE, REG-FORCE, and NUDGE experiments, respectively. Specifically, the red line in (c) denotes the regressed TAUXA-N4 in response to the SSTA in the FULL-FORCE run. The values in the CORR array, arranged from left to right, stand for the correlation coefficients between observation and the three experiments from 1982 to 2020 (39 years). The values in the RMSE array, arranged from left to right, denote the corresponding root mean square errors of observation and three experiments from 1982 to 2020.

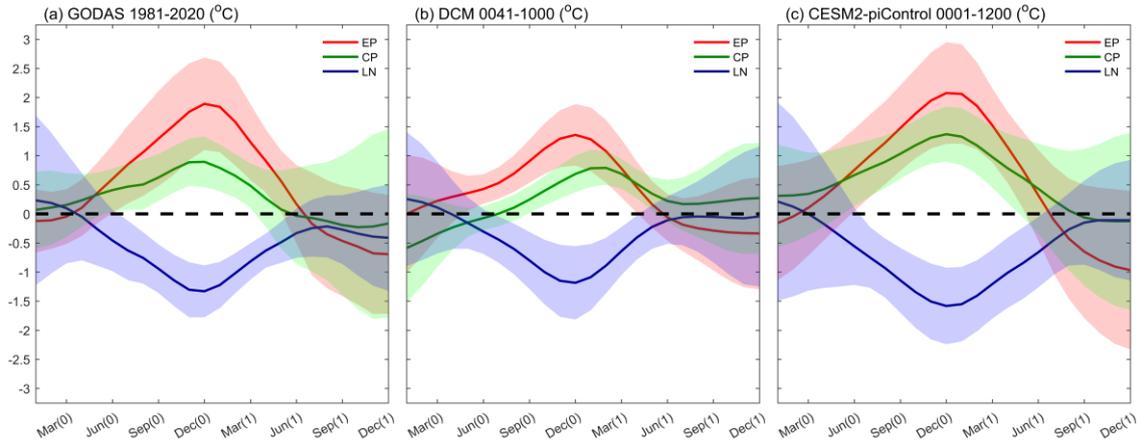

**Supplementary Fig. S2.** The composite evolution of Niño 3.4 region averaged SSTA (°C) for all the EP El Niño (EP), CP El Niño (CP), and La Niña (LN) for (a) GODAS 1981-2020, (b) DCM 0041-1000, (c) CESM2-piControl 0001-1200. The solid line and shading denote the mean value and standard deviations of all the corresponding events. The horizontal labels indicate the end of the corresponding months for the development year (0) and the subsequent year (1).

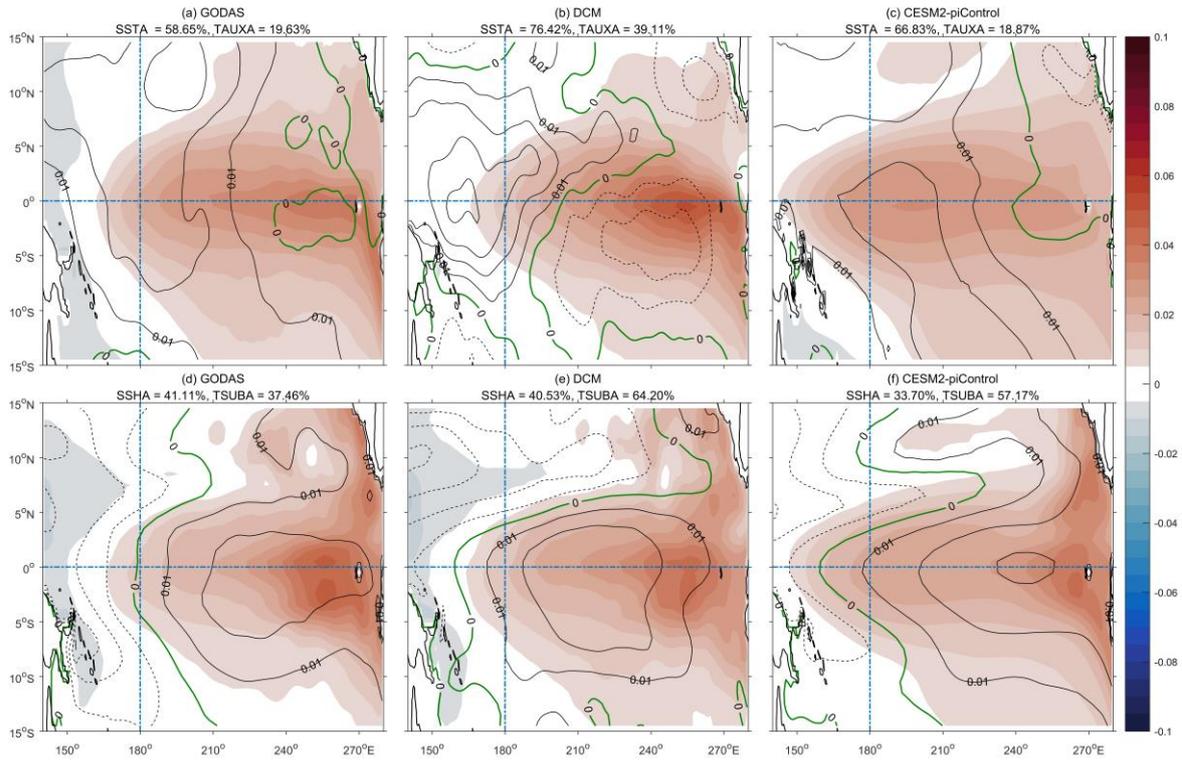

**Supplementary Fig. S3.** The normalized first-leading empirical orthogonal functions (EOF1) of SSTA (a-c, shading), TSUBA (d-f, shading), TAUXA (a-c, contour), SSHA (d-f, shading) for GODAS (1981-2023), DCM (0041-0190), and CESM2-piControl (0001-0150). The shading interval is 0.005 for SSTA and TSUBA. The equidistant contour interval is 0.01 for TAUXA and SSHA. The solid black, solid green, and dashed black contour lines denote positive, zero, and negative values, respectively. The values in the subtitles represent the explained variance of EOF1 for the corresponding variable. The EOF region (15°S-15°N, 140°E-280°E) follows Cai et al. (2018).

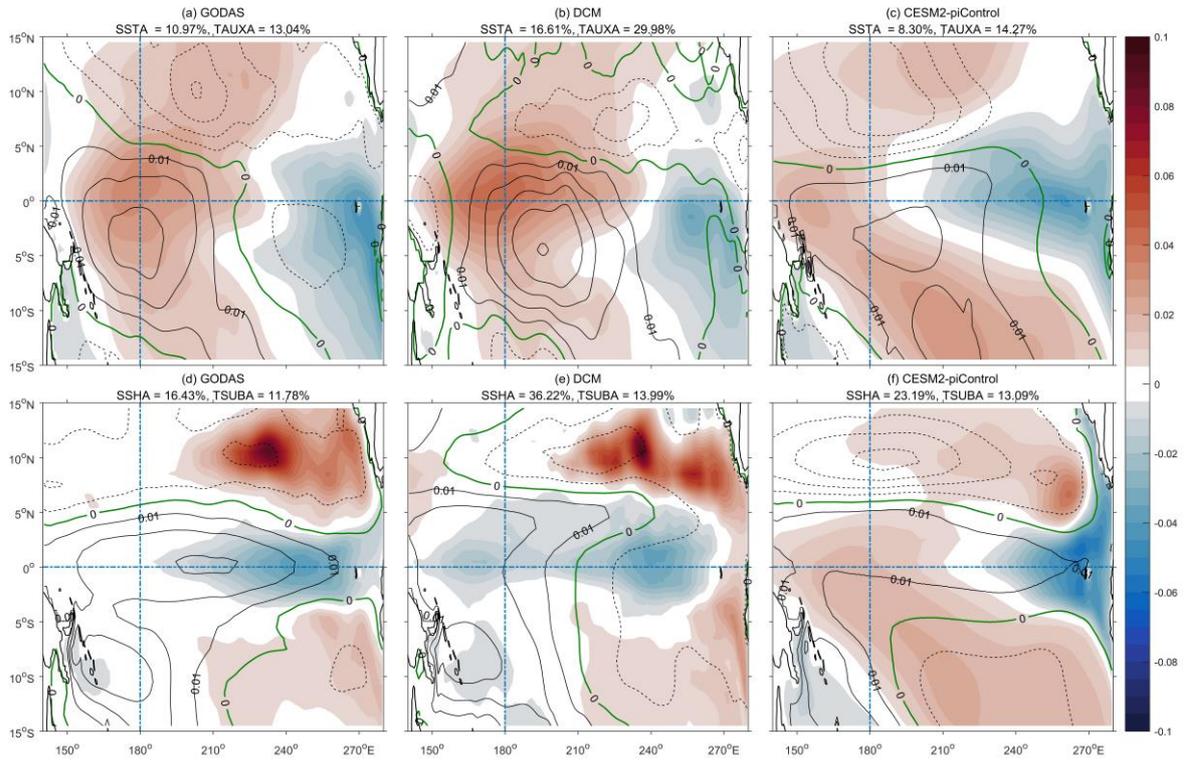

**Supplementary Fig. S4.** Same as Figure S3, but for the normalized second-leading empirical orthogonal function (EOF2).

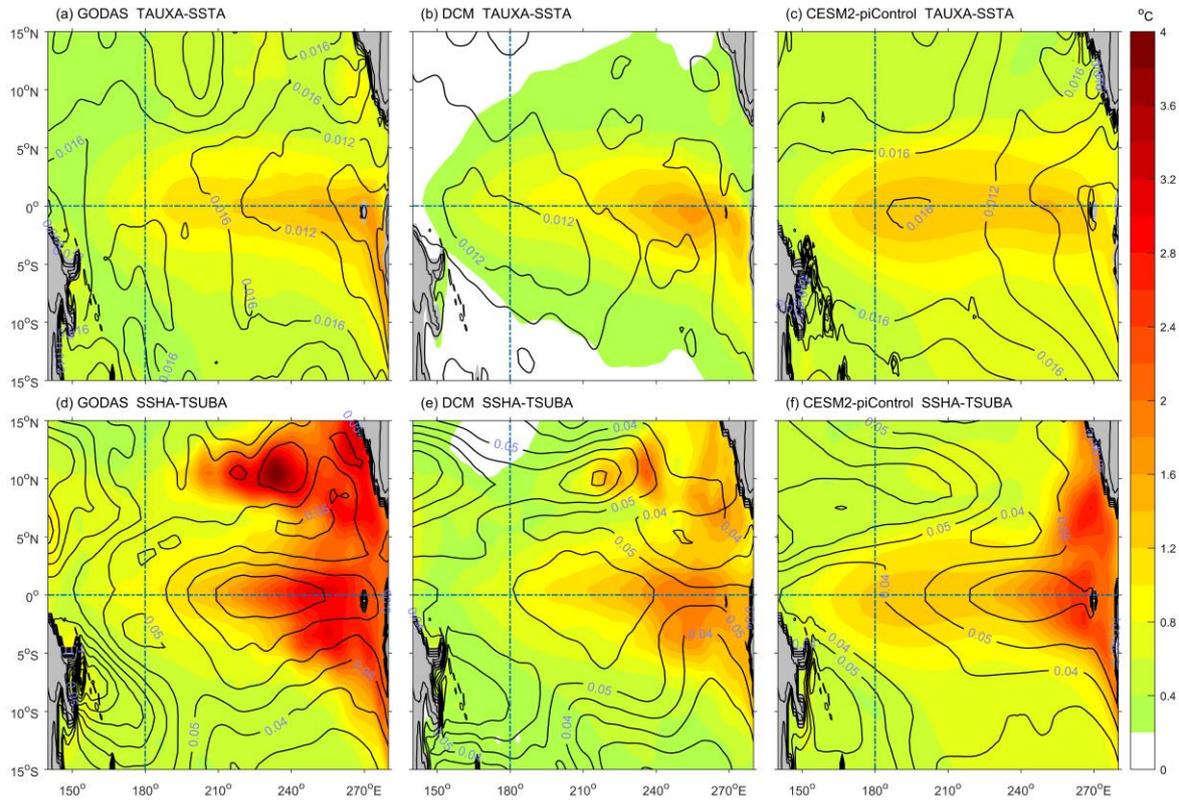

**Supplementary Fig. S5.** The all-year standard deviations of SSTA (a-c, shading), TSUBA (d-f, shading), TAUXA (a-c, contour), SSHA (d-f, shading) for GODAS (1981-2023), DCM (0041-1000), and CESM2-piControl (0001-1200). The shading interval is 0.2°C for SSTA and TSUBA. The contour interval is 0.004 N m$^{-2}$ and 0.01 m for TAUXA and SSHA, respectively.

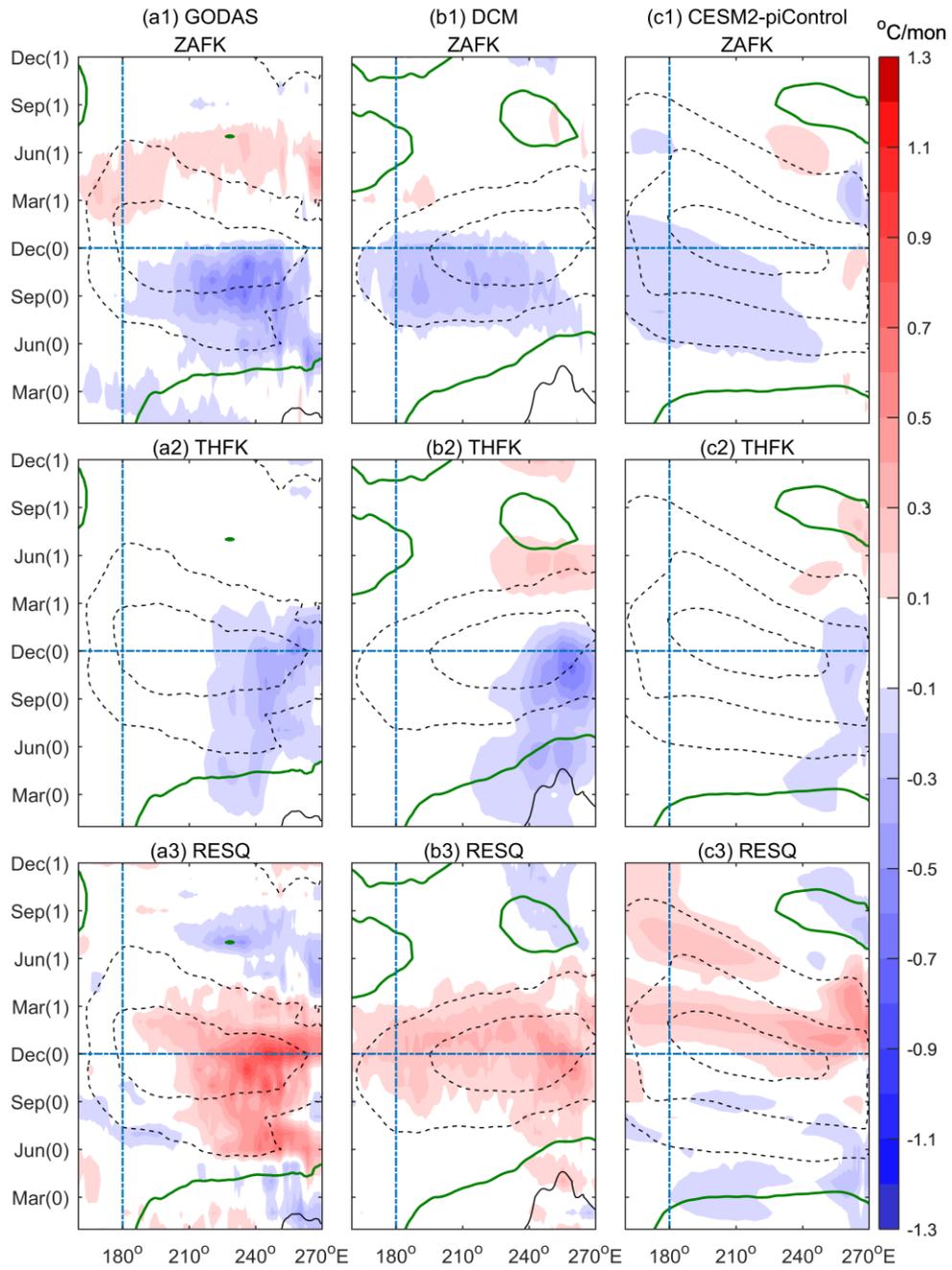

**Supplementary Fig. S6.** Same as Fig. 4 in the main body, but for La Niña.

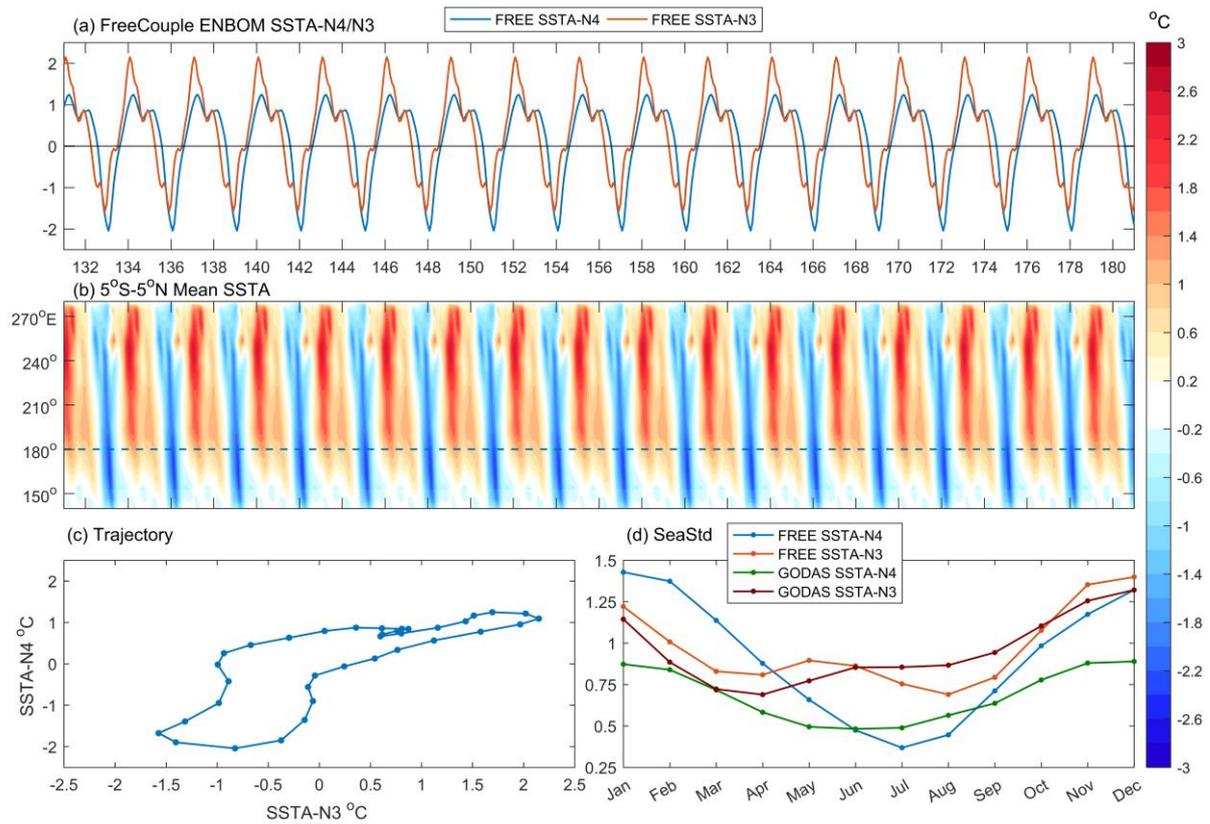

**Supplementary Fig. S7.** The time evolution of (a) SSTA-N4 (solid blue line), SSTA-N3 (solid orange line), and (b) 5°S-5°N mean SSTA (shading) for the free-coupled ENBOM itself (indicated by FREE) without the nudging of CF23. The shading interval is 0.2°C. The horizontal ticks of (a) and (b) stand for the start of model years. (c) shows the trajectory of SSTA-N4 and SSTA-N3 during the model year 0041-0200. (d) The solid blue and solid orange lines denote the seasonal standard deviations (°C) of SSTA-N4 and SSTA-N3 for the free-coupled ENBOM during the model year 0041-0200 (FREE SSTA-N4 and FREE SSTA-N3), respectively. The solid green and purple lines represent the standard deviations of SSTA-N4 and SSTA-N3 for the 1981-2020 GODAS dataset (GODAS SSTA-N4 and GODAS SSTA-N3), respectively.

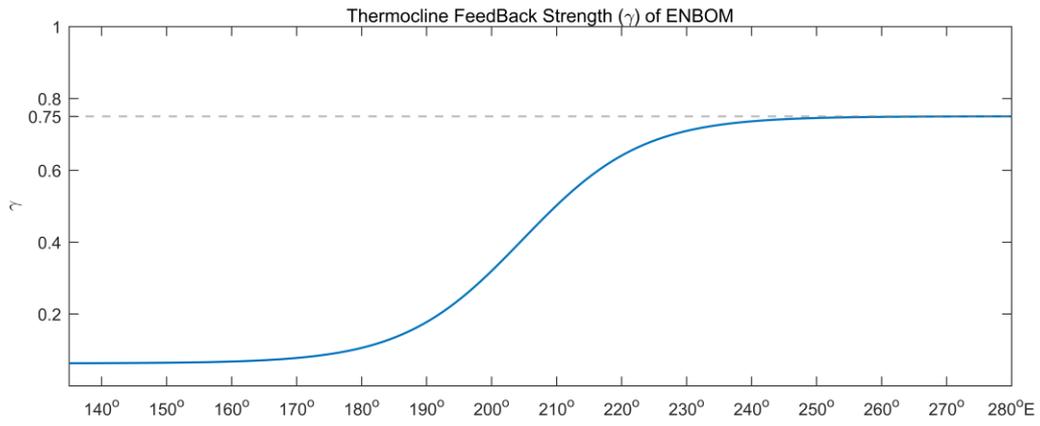

**Supplementary Fig. S8.** The zonal profile of thermocline feedback strength $\gamma$ with a maximum value of 0.75 for ENBOM.